\documentclass[12pt]{article}
\usepackage{graphicx}
\usepackage{amsmath,amssymb,amsfonts}

\begin{document}

\title{Effects of gravitational waves on the polarization of pulsars}

\author{Shahen Hacyan
}

\renewcommand{\theequation}{\arabic{section}.\arabic{equation}}

\maketitle
\begin{center}

{\it  Instituto de F\'{\i}sica,} {\it Universidad Nacional Aut\'onoma de M\'exico,}

{\it A. P. 20-364, M\'exico D. F. 01000, Mexico.}

\end{center}
\vskip0.5cm

\begin{abstract}

The polarization of electromagnetic waves in the presence of a gravitational  wave is analyzed. The rotation of
the polarization angle and the Stokes parameters are deduced. A possible application to the detection of
stochastic background of gravitational waves is proposed as a complement to the pulsar timing method.

\end{abstract}


\vskip3.5cm
e-mail: hacyan@fisica.unam.mx

%Key words: .

 \maketitle

\newpage

\section{Introduction}

Pulsar timing is a promising method for the detection  of stochastic background of gravitational waves (GWs) (see,
e.g., Hobbs  et al. \cite {hobbs}, Wang et al. \cite {wang}). As shown by Hellings and Downs \cite{hd}, the shift
in the pulses arrival time can be related to the spectrum of this background. Pulsars also exhibit, besides very
regular pulses, important properties such a the polarization of their signals. This polarization has a complicated
structure (see,  e.g., Craig \cite {cra}), but it can be affected by GWs and it may be possible therefore to
extract some information about their presence in the galaxy. As shown in the present paper, this could be achieved
by adapting and complementing the statistical analysis proposed by Hellings and Downs \cite{hd}.

The aim of  this paper is to analyze  the interaction of electromagnetic waves  with a plane fronted GW, and to
deduce the corresponding formulas for the Stokes parameters and the rotation angle of the polarization. A similar
problem was studied from a formal point of view in a previous paper \cite{h1}, with the GW described by the
Ehlers-Kundt metric \cite{ek}, which is an exact solution of the Einstein equations. It was shown that there is a
rotation of the polarization angle in phase with the GW. In the present paper, we apply a similar analysis to a GW
described in the standard form with its two independent polarization states. Section 2 is devoted to the
mathematical formalism in the short wave-length approximation, and the electromagnetic coherency matrix is
obtained in Section 3. A possible application to the detection of stochastic GWs background is discussed in
Section 4.

\section{The electromagnetic field}

The metric of a plane gravitational wave in the weak field limit is
\begin{equation}
ds^2 = -2 du ~dv+ (1+f)dx^2 + (1-f)dy^2 +2 g~ dx~dy .\label{1}
\end{equation}
The relation with Minkowski coordinates $t$ and $z$ is
$$
u= \frac{1}{\sqrt{2}}(t-z) ,~~ \quad  v= \frac{1}{\sqrt{2}}(t+z),
$$
and the two degrees of polarization of the GW are given by the potentials $f(u)$ and $g(u)$, which are functions
of $u$ only. In the following, quadratic and higher terms in  $f$ and $g$ are neglected, coordinates are taken in
the order $(u,v,x,y)$,  and we set $c=1$.

 In the short wave-length approximation (see, {\it e.g.},
Misner, Thorne and Wheeler \cite{mtw}  or Ref. \cite{h1}), the
electromagnetic potential four-vector $A_{\alpha}$ is written in the form
\begin{equation}
A_{\alpha}=a_{\alpha} e^{iS},
\end{equation}
in terms of an auxiliary four-vector $a_{\alpha}$ and the eikonal function $S$ which satisfies the equation
\begin{equation}
g^{\mu \nu} S_{,\mu}S_{,\nu}=0.
\end{equation}
Defining the null four-vector
$$K_{\alpha} = S_{,\alpha},$$
the Maxwell equations can be expanded in descending powers of $S$. The first term in the expansion implies
\begin{equation}
a^{\alpha} K_{\alpha} =0
\end{equation}
and the electromagnetic field turns out to be
\begin{equation}
F_{\alpha \beta} = i(K_{\alpha} a_{\beta} - K_{\beta} a_{\alpha})e^{i S}.
\end{equation}

The second term in the expansion implies
\begin{equation}
K^{\beta} (a_{\alpha ,\beta} - a_{\beta ,\alpha})+ (a_{\alpha} K^{\beta} -  a^{\beta} K_{\alpha} )_{;\beta}=0.
\end{equation}

Choosing a gauge $A_v=0$ as if Ref. \cite{h1}, a particular solution of the above equations turns out to be
\begin{eqnarray}
a_{u} &=& -\frac{1}{k_v} \Big[k_x a_x + k_y a_y \Big] \\ \nonumber
a_{v}&=&0 \\ \nonumber
a_{x} &=& \Big(1+ \frac{1}{2}f(u)\Big)~\overline{a}_x +\frac{1}{2}g(u) ~\overline{a}_y \\
\nonumber a_{y} &=& \frac{1}{2}g(u) ~\overline{a}_x +\Big(1- \frac{1}{2}f(u)\Big)~\overline{a}_y  \label{aaaa}
\end{eqnarray}
where $\overline{a}_x$ and $\overline{a}_y$ are constants.

As for the null-vector $K_{\alpha}$, its components are
\begin{equation}
K_{\alpha}= \Big(k_u [1- {\cal F}(u)]~,~ k_v~,~ k_x~,~ k_y \Big),
\end{equation}
where
\begin{equation}
{\cal F}(u) = f(u) \cos 2\phi + g(u) \sin 2\phi ,
\end{equation}
$k^{\alpha}= ( \omega ,{\bf k})$ is the wave-vector in Minkowski coordinates, and the angle $\phi$ is defined by $k_x =
k_{\bot} \cos \phi$ and $k_y = k_{\bot} \sin \phi$, with $k_{\bot}= k_x^2 + k_y^2$. In particular
$$
k_u = - \frac{\omega + k_z}{\sqrt{2}}
$$
$$
k_v = - \frac{\omega - k_z}{\sqrt{2}}.
$$

(In the limit of flat space-time, the above solution describes a plane electromagnetic wave with the electric
field
\begin{equation}
{\bf E} =i(\omega {\bf \overline{a}} - \overline{a}_z {\bf k})e^{i\bar{S}},
\end{equation}
where $\bar{S} =k_{\alpha} x^{\alpha}=-\omega t +{\bf k}\cdot {\bf r}$ and $\omega \overline{a}_z = {\bf k}\cdot
\overline{{\bf a}}$ (see the Appendix of Ref. \cite{h1}).)

We can now choose a time-like unitary vector
\begin{equation}
t^{\alpha}= \frac{1}{\sqrt{2}} \Big(1,1,0,0 \Big)
\end{equation}
which is also a geodesic, that is, it is tangent to the world-line of an inertial observer fixed at a point with
constant coordinates $(x,y,z)$.

The observed frequency of the electromagnetic wave is defined covariantly as
$$
\Omega (u)= - K_{\alpha} t^{\alpha} = -\frac{1}{\sqrt{2}} (K_u+k_v)=
$$
\begin{equation}
\omega [1 - \frac{1}{2}(1 + \cos \theta ) {\cal F}(u)],
\label{om}
\end{equation}
where $\theta$ is the angle between ${\bf k}$ and the direction of propagation of the GW, and of course $\omega =  - k_{\alpha} t^{\alpha}$, which is the frequency observed in the absence of the
GW. From this last formula, it follows that the Doppler shift for a pulsar located at a
distance $l$ from Earth and at an angle $\theta$ with respect to the direction of propagation of the GW is
$$
[\Omega(t)-\Omega(t-l-l\cos \theta)]/\omega,
$$
which is the formula deduced by Estabrook and Wahlquist \cite{ew} for the pulse frequency of a pulsar (for $f$
type polarization only).

\section{Coherency matrix}

Given the time-like vector $t^{\alpha}$, the electric field can be defined as $E^{\alpha} \equiv F^{\alpha \beta}
t_{\beta}$, where $F^{\alpha \beta}$ is the electromagnetic tensor. In the short wave-length approximation
\cite{h1}
\begin{equation}
E_{\alpha} =i[\Omega a_{\alpha}+ (a_{\beta} t^{\beta})K_{\alpha}] e^{iS}.
\end{equation}

Two space-like unitary vector $e^{(i)}_{\alpha}$ ($i=1,2$), orthogonal to each other and to both $t^{\alpha}$ and
$K_{\alpha}$, can be defined. Accordingly
\begin{equation} E_{\alpha} e^{(i)\alpha}=i\Omega a_{\alpha}
e^{(i)\alpha}e^{iS}.
\end{equation}

A possible choice satisfying all the orthogonality conditions is
\begin{equation}
e^{(1)\alpha}= \frac{1}{k_{\bot}\sqrt{1-{\cal F}} } \Big(0,0,-k_y,k_x \Big)
\end{equation}
and
$$
e^{(2)}_{\alpha}= - \frac{k_{\bot}\sqrt{1-{\cal F}}}{K_u+k_v } \Big(-1,1,0,0 \Big)-
$$
\begin{equation}
\frac{1}{k_{\bot} \sqrt{1-{\cal F} }}~ \frac{(K_u - k_v )}{K_u +k_v } \Big(0,0,k_x,k_y \Big).
\end{equation}
Accordingly, we have
\begin{equation}
a_{\alpha} e^{(1)\alpha}= \frac{1}{k_{\bot}\sqrt{1-{\cal F}} } (-a_x k_y + a_y k_x )
\end{equation}
\begin{equation}
a_{\alpha} e^{(2)\alpha}= \frac{a_u k_v}{k_{\bot}\sqrt{1-{\cal F}} }
\end{equation}
and therefore
\begin{equation}
E^{(1)} \equiv E_{\alpha} e^{(1)\alpha}= i\frac{\Omega }{k_{\bot} \sqrt{1-{\cal F}}}(a_y k_x -a_x k_y )e^{iS}
\end{equation}
\begin{equation}
E^{(2)} \equiv E_{\alpha} e^{(2)\alpha}= i\frac{\Omega }{k_{\bot} \sqrt{1-{\cal F}}} a_u k_v e^{iS}.
\end{equation}

These formulas can be written in the compact form
\begin{equation}
\begin{pmatrix}
  E^{(1)}\\
  E^{(2)}
\end{pmatrix}
= i \frac{\Omega e^{iS}}{k_{\bot}}
 \Big[  \mathbf{1} + \frac{1}{2} {\cal G}
\begin{pmatrix}
  0 & 1 \\
  -1 & 0
\end{pmatrix}
\Big]
\begin{pmatrix}
  -k_y & k_x \\
  k_x & k_y
\end{pmatrix}
\begin{pmatrix}
  \overline{a}_x \\
  \overline{a}_y
\end{pmatrix},
\end{equation}
where
\begin{equation}
{\cal G}(u) =g(u) \cos 2\phi -f(u) \sin 2\phi,
\end{equation}
or even more compactly as
\begin{equation}
\begin{pmatrix}
  E^{(1)}\\
  E^{(2)}
\end{pmatrix}
= i \frac{\Omega }{\omega}e^{i(S-\bar{S}) }
 \Big[  \mathbf{1} + \frac{1}{2} {\cal G}
\begin{pmatrix}
  0 & 1 \\
  -1 & 0
\end{pmatrix}
\Big]
\begin{pmatrix}
  \bar{E}^{(1)}\\
  \bar{E}^{(2)}
\end{pmatrix},\label{rot}
\end{equation}
where $\bar{E}^{(i)}$ are the values of the electric field in the absence of the GW. Thus the contribution of the
GW is made implicit through the functions ${\cal F}(u)$ and ${\cal G}(u)$, and it follows from this last formula
that the GW produces a rotation of the polarization vector by an angle $-(1/2) {\cal G}(u)$.

The coherency matrix is
\begin{equation}
\mathbb{S}\equiv
\left(
  \begin{array}{c}
    E^{(1)} \\
    E^{(2)} \\
  \end{array}
\right)
\left(
  \begin{array}{cc}
    E^{(1)*}, & E^{(2)*} \\
  \end{array}
\right)=
\begin{pmatrix}
  S_0+S_1 & S_2-iS_3 \\
  S_2+iS_3 & S_0-S_1
\end{pmatrix} ,
\end{equation}
where $S_i$ are the Stokes parameters (in the notation of Born and Wolf \cite{bw}). It then follows with some
algebraic manipulations (such as in Ref. \cite{h1}) that
\begin{equation}
\mathbb{S} = \frac{\Omega^2}{\omega^2} \overline{\mathbb{S}} - {\cal G}
\begin{pmatrix}
  -\bar{S}_2 & \bar{S}_1 \\
  \bar{S}_1 & \bar{S}_2
\end{pmatrix}
,
\end{equation}
where $\overline{\mathbb{S}}$ is the coherency matrix and $\bar{S}_i$  the Stokes parameters in the absence of the
GW.

More explicitly
\begin{eqnarray}
S_0 &=&  \bar{S}_0 [1 - (1 + \cos \theta ) {\cal F}] \\ \nonumber
S_1 + i S_2 &=& (\bar{S}_1 + i\bar{S}_2)[1 -(1 + \cos \theta ) {\cal F} -i {\cal G} ] \\ \nonumber
S_3 &=&  \bar{S}_3 [1 - (1 + \cos \theta ) {\cal F}].\label{fin}
\end{eqnarray}

It can be seen from these last formulas that if the light is not initially polarized, namely $<\bar{S}_i>=0$ for
$i=1,2,3$, it will not be polarized by the GW. However, if the light is originally polarized, the angle of the
polarization ellipse oscillates at the passage of the GW. This can be seen by setting
$$
S_1 + iS_2 = S_0 \cos 2\chi ~ e^{2i \psi}
$$
$$
S_3 = S_0 \sin 2\chi ,
$$
in terms of the Poincar\'e angles (see, e.g. \cite{bw}): $\psi$ specifies the orientation of the polarization
ellipse and $\chi$ its ellipticity. It thus follows  that the ellipticity of the polarized light is not altered by
a GW, but the polarization angle $\psi$ changes as
\begin{equation}
\psi \rightarrow \psi - \frac{1}{2} {\cal G}(u),\label{psi}
\end{equation}
as already implied by \eqref{rot}.

The intensity of the wave changes as
$S_0$  and the variation of the frequency is given by \eqref{om}.

\section{Correlations}

It is  possible to adapt the analysis of Hellings and Downs \cite{hd} to the polarization shift as given by the
above formulas. The change in the polarization angle $\Delta \psi$ observed on pulsar number $i$ is
\begin{equation}
\Delta \psi_i = -\alpha_i g(t)  + \beta_i f(t)  + n_i (t),
\end{equation}
where $f(t)$ and $g(t)$ are the GW signals common to all pulsars,
$$
\alpha_i = \frac{1}{2} \cos 2\phi_i,
$$
$$
\beta_i = \frac{1}{2}  \sin 2\phi_i,
$$
are angles factors for the $i$ th pulsar, and $n_i (t)$ represents all noise sources unique to the pulsar,
including terms depending on its position.

%An entirely similar formula applies to the change in the intensity $I$ of the electromagnetic wave:
%\begin{equation}
%\frac{\Delta I_i }{I_i} = \alpha_i f(t)  + \beta_i g(t)   + n'_i (t),
%\end{equation}
%with the same coefficients $ \alpha_i$ and $\beta_i$, and a different noise source $n'_i$.

The corresponding analysis is just as in Hellings and Downs \cite{hd}. The correlation $<\Delta \psi_i \Delta
\psi_j>$ for two pulsars is
\begin{equation}
{\cal C}_{ij} = \alpha_i \alpha_j <f^2> + \beta_i \beta_j <g^2> ,
\end{equation}
in terms of the correlations of the potential functions, the basic assumptions being that the processes $n_i$,
$n_j$, together with $f(t)$ and $g(t)$, are all uncorrelated among themselves.

For an isotropic background of GWs, one has to calculate the angle averages for pairs of pulsars
\begin{equation}
\alpha_{ij} \equiv \frac{1}{4\pi} \int \alpha_i \alpha_j ~d\Omega = \frac{1}{4\pi} \int \beta_i \beta_j ~d\Omega.
\end{equation}
The result is
$$
\alpha_{ij}= \frac{1}{8} + \frac{1}{4} (1-\cos \gamma_{ij}) {\rm ln}\Big(\frac{1-\cos \gamma_{ij}}{2}\Big)+
$$
\begin{equation}
~~~~~\frac{1}{4} (1+\cos \gamma_{ij}) {\rm ln}\Big(\frac{1+\cos \gamma_{ij}}{2}\Big),\label{aij}
\end{equation}
where $\gamma_{ij}$ is the angle between the two pulsars (see the Appendix for an outline of the derivation).

The rest of the formulation is \textit{mutatis mutandis} just as in Ref. \cite{hd} and will not be repeated
here.

\section{Conclusion}

The influence of a GW on the polarization of light is made explicit by the formulas obtained above. The basic
effect of a GW is to produce an oscillation of the electromagnetic wave polarization. This effect could be applied
to the detection of GWs. An example of such an application was outlined as a complement to the pulsar timing
method.

\section*{Appendix}

Consider first gravitational waves propagating along a particular direction, say the $z$ axis. An average of the
product $\alpha_1 \alpha_2$ over the azimuthal angle $\phi$ is
$$
\frac{1}{8\pi}\int_0^{2\pi} d\phi ~ \cos (2(\phi_1 - \phi)) \cos (2(\phi_2 - \phi))= \frac{1}{8} \cos (2(\phi_1
-\phi_2)),
$$
where $\phi_i$ are the azimuthal angles of the pulsars. The same result applies to the averaged product $\beta_1
\beta_2$. The above expression can be written in a coordinate independent form as
$$
  \frac{[\cos \gamma - ({\bf a}_1 \cdot {\bf g})({\bf a}_2 \cdot {\bf
g})]^2}{4[1  - ({\bf a}_1 \cdot {\bf g})^2]  [1-  ({\bf a}_2 \cdot {\bf g})^2]} - \frac{1}{8},
$$
where ${\bf a}_i$ are the unit vectors in the directions of the pulsars and ${\bf g}$ is the unit vector in the
direction of the considered GWs.

If $\gamma$ is the angle between the two pulsars, we can now choose without losing generality:
$$
{\bf a}_1 = (\cos (\gamma /2), \sin (\gamma /2), 0 )
$$
$$
{\bf a}_2= (\cos (\gamma /2), -\sin (\gamma /2), 0 ),
$$
and it then follows that an integration over all the directions ${\bf g}$ yields
$$
\alpha_{12} = \frac{1}{16 \pi}\int d\Omega_{{\bf g}}~ \Big[ \frac{(\cos \gamma - g_1
g_2)^2}{(1-g_1^2)(1-g_2^2)}-\frac{1}{2} \Big],
$$
where
$$
g_1 = \sin \theta \cos (\phi - \gamma/2)
$$
$$
g_2 = \sin \theta \cos (\phi + \gamma/2)
$$

Using as an intermediary step the general formula
$$
\int_0^{2\pi} d\phi ~ \frac{\Big[\cos \gamma - (1/2) \sin^2 \theta (\cos (2\phi) +\cos \gamma)\Big]^2}{[1 - \sin^2
\theta \cos^2(\phi + \gamma/2)]  [1 - \sin^2 \theta \cos^2 (\phi - \gamma/2)]}=
$$
$$
2\pi \Big[ 1 - 2  ~\frac{|\cos \theta| ~(1+ \cos^2 \theta)}{\sin^4 \theta +4 \cos^2 \theta /\sin^2\gamma } \Big],
$$
one obtains Eq. \eqref{aij}.

\end{document}